# Observation of higher-order solitons in defocusing waveguide arrays

Eugene Smirnov, Christian E. Rüter, and Detlef Kip

*Institute of Physics and Physical Technologies, Clausthal University of Technology,*

*38678 Clausthal-Zellerfeld, Germany*

Yaroslav V. Kartashov and Lluis Torner

*ICFO-Institut de Ciencies Fotoniques, Mediterranean Technology Park, and*

*Universitat Politecnica de Catalunya, 08860 Castelldefels (Barcelona), Spain*

We observe experimentally higher-order solitons in waveguide arrays with defocusing saturable nonlinearity. Such solitons can comprise several in-phase bright spots and are stable above a critical power threshold. We elucidate the impact of the nonlinearity saturation on the domains of existence and stability of the observed complex soliton states.

Transverse periodic refractive index modulations strongly affect light propagation and, in combination with nonlinearity, they can result in the formation of discrete and lattice solitons encountered in various settings [1], including waveguide arrays imprinted in semiconductors [2,3], photonic lattices [4-7], and arrays made in photovoltaic crystals [8,9]. Importantly, the latter exhibit saturable nonlinearities that strongly impact lattice soliton properties. Thus, nonlinearity saturation results in the swapping of the stability properties between different soliton families [10] and radiationless propagation of tilted beams across the lattice [11,12]. Periodic lattices support gap solitons [3,4,7-9], dark and anti-dark solitons [13], even and twisted solitons [14], or higher-order soliton trains [10,15,16]. Higher-order solitons have been also studied in two-dimensional lattices [17-20]. While the existence of higher-order lattice solitons in *focusing* media was confirmed in experiments, higher-order solitons in *defocusing* lattices have never been observed to date.

In this Letter we report on the experimental observation of higher-order solitons in a waveguide array imprinted in a LiNbO$_3$ crystal with *defocusing saturable nonlinearity*. Such higher-order solitons comprise several in-phase bright spots and we found that they are stable at high enough power levels. We show that nonlinearity saturation strongly impacts the domains of soliton existence and stability.

To elucidate the conditions for higher-order soliton formation in defocusing lattices we explore their properties theoretically. We consider the nonlinear Schrödinger equation for a dimensionless field amplitude $q$ describing beam propagation in a defocusing saturable medium:

$$i\frac{\partial q}{\partial \xi} = -\frac{1}{2}\frac{\partial^2 q}{\partial \eta^2} + \frac{q|q|^2}{1+S|q|^2} - pRq. \quad (1)$$

Here the longitudinal $\xi$ and transverse $\eta$ coordinates are normalized to diffraction length and beam width, respectively, $S$ is the saturation parameter, $p$ is the lattice depth, and the function $R(\eta) = \cos^2(\Omega\eta)$ describes the refractive index profile. The values of parameters in Eq. (1) were set in accordance with the actual data of our sample. Thus, for a beam width $\sim 5\,\mu\text{m}$ and wavelength $\lambda = 532\,\text{nm}$, a lattice period $7.6\,\mu\text{m}$ corresponds to $\Omega = 2.067$, a lattice depth $p \approx 17$ corresponds to a refractive index variation $\delta n \approx 0.0022$, a crystal length $25\,\text{mm}$ corresponds to $\xi = 38$, and the maximal nonlinear contribution to the refractive index due to photovoltaic effect $\delta n_{\text{nl}} \sim 2 \times 10^{-4}$ yields $S \approx 0.5$. For the sake of generality, we also consider other $p$ and $S$ values, keeping $\Omega$ fixed. Equation (1) conserves the energy flow $U = \int_{-\infty}^{\infty} |q(\eta,\xi)|^2 \, d\eta$.

We search for soliton solutions of Eq. (1) in the form $q = w(\eta)\exp(ib\xi)$, where $b$ is the propagation constant. Soliton solutions are encountered for propagation constants falling into finite gaps of the lattice spectrum. Here we focus only on the first finite gap where lowest-order odd soliton acquires its intensity maximum in the lattice maximum. The lattice can also support complex soliton structures that might be intuitively viewed as nonlinear combinations of odd solitons residing in adjacent channels. The simplest even solitons of this type resembling in-phase combinations of odd solitons are shown in Figs. 1(a) and 1(b). Expansion across the lattice for $b$ values in the vicinity of the gap edges [Fig. 1(a)] is replaced by a strong soliton localization into two bright spots in the middle of the gap [Fig. 1(b)]. The stronger the lattice the higher the soliton localization in the gap depth. For large enough $S$ the energy flow of even solitons is a monotonically decreasing function of $b$ everywhere, except for narrow regions close to the upper and lower cut-offs for existence [not visible in Fig. 1(c)] where $dU/db > 0$.

In contrast to the cubic case, in saturable media the domain of existence of even soliton does not occupy the whole gap. At fixed $p$ the lower cut-off $b_{\text{low}}$ coincides with the lower gap edge, however, when the saturation parameter exceeds a certain minimal value ($S_{\text{m}} \approx 0.12$ for $p = 17$) the cut-off $b_{\text{low}}$ increases with $S$. Similarly, the upper cut-off $b_{\text{upp}}$ amounts always to smaller values than the upper gap edge, but approaches it as $S \to 0$. With increasing $S$ the cut-off $b_{\text{upp}}$ decreases, so that at a certain critical value of $S$ ($S_{\text{cr}} \approx 1.74$ for $p = 17$) the domain of existence of even solitons shrinks (Fig. 1(d)). This is in contrast to the existence domain for usual odd solitons that never vanishes with increasing $S$. The width of the existence domain for even solitons in the plane $(p,b)$ decreases with decreasing $p$, and shrinks completely when $p$ decreases below a critical value ($p_{\text{cr}} \approx 7.23$ for $S = 0.5$). Thus, too shallow lattices cannot support even solitons. In deep lattices the upper cut-off $b_{\text{upp}}$ gradually approaches the upper gap edge, while $b_{\text{low}} \approx b_{\text{upp}} - 1/S$. Soliton localization in the vicinity of cut-offs is determined by the location of $b_{\text{low}}$ or $b_{\text{upp}}$ inside the gap.

Linear stability analysis indicates that even solitons are stable for high enough energy flows, when $b_{\text{low}} < b < b_{\text{cr}}$. The critical propagation constant for stabilization is depicted by the red (dashed) curve in Fig. 1(d). We found only a domain of oscillatory instability near the upper cut-off for existence (at $b_{\text{cr}} < b < b_{\text{upp}}$), where a perturbation may cause the decay of an even soliton into an odd one, and very narrow domains of exponential instability in the regions adjacent to cut-offs where $dU/db \geq 0$. In Fig. 1(f) we plot the dependence of real part $\delta_r$ of perturbation growth rate $\delta = \delta_r + i\delta_i$ on propagation constant where one can clearly see that $\delta_r$ vanishes at $b = b_{\text{cr}}$. The maximal growth rate decreases with $S$, so that close to $S_{\text{cr}}$ the domain of instability vanishes. Importantly, stable combinations of multiple in-phase odd solitons are possible, too, which feature properties similar to that of even solitons. Thus, defocusing optical lattices allow the formation of stable extended soliton trains.

To confirm this prediction experimentally, we used a 25 mm long waveguide array with period 7.6 $\mu$m fabricated in copper-doped $LiNbO_3$. The fabrication procedure of such arrays, consisting of about 250 channels, has been described in detail elsewhere [8]. In our setup, a CW frequency-doubled $YVO_4$ laser, $\lambda = 532$ nm, illuminates an amplitude mask containing two circular holes, which are imaged onto the sample input facet in order to excite two adjacent in-phase channels of the array. Alternatively, a single focused Gaussian beam can be used for simultaneous in-phase excitation of two or three channels, where the width of this beam is adjusted with the help of a cylindrical lens in front of the microscope lens used for incoupling. The output patterns on the sample end facet are imaged onto a CCD camera.

In Fig. 2(a) linear discrete diffraction of light from a single excited channel is monitored, from which we can estimate a coupling length of $L_c \approx 4.5$ mm in our sample. Even solitons and higher-order soliton trains form when amplitude and width of in-phase beams launched into neighboring channels are properly adjusted. An illustrative example of the observed images from the sample end facet are shown in Fig. 2(b) where the amplitude mask was employed to excite two in-phase beams with incoupled power of 10 $\mu$W. The observations are in agreement with simulations for Gaussian beams, i.e., $A\exp[-(\eta/W)^2]$. Alternatively, even solitons may also be excited with a single beam of the same optical power that is launched between two lattice channels [Fig. 2(c)], while higher-order trains can be excited using a broader beam that covers several channels [Fig. 2(d)]. This is consistent with numerical simulations, which in this case were conducted with super-Gaussian inputs $A\exp[-(\eta/W)^4]$. In all cases, the simulations indicate that soliton excitation is efficient and occurs with short sample lengths.

We also investigated the temporal dynamics of even soliton formation (Fig. 3). The linear output pattern in Fig. 3(a) is a superposition of the discrete diffraction [see Fig. 2(a)] of two single-channel excitations. In Fig. 3(b) the build-up of an even soliton for excitation with two in-phase Gaussian beams is displayed. Initially, asymmetries in the output pattern during build-up of the nonlinearity are observed, which we attribute to fragility of soliton trapping in the vicinity of low-power cut-off. Note that during this transient stage, higher input powers would be required for stable propagation of even excitations than in the steady-state regime, when the nonlinearity strength is higher. As a result, for a fixed input power one observes the transition between stages of linear diffraction, unstable propagation, and stable propagation upon temporal build-up of nonlinearity. In Fig. 3(c) we have blocked one of the input beams after formation of the even soliton, and the temporal dynamics shows a conversion of the output pattern into an odd soliton. This conversion is accompanied by transient energy oscillation among the two channels.

In summary, we have observed experimentally and analyzed theoretically higher-order solitons in one-dimensional waveguide arrays in defocusing saturable nonlinear media. Our observations highlight the important effects afforded by nonlinearity saturation, and confirm the possibility of packing of several lattice solitons into compact stable trains.

# References with titles

# References without titles

# Figure captions

Figure 1 (color online). Profiles of even solitons with $b = 11.34$ (a) and $b = 10$ (b) at $p = 17$ and $S = 0.5$. (c) Energy flow versus propagation constant at $p = 17$. Domains of existence for even solitons on $(S,b)$ plane at $p = 17$ (d) and on $(p,b)$ plane at $S = 0.5$ (e). Red (dashed) curve in panel (d) indicates the upper border of stability domain. Real part of perturbation growth rate versus propagation constant at $p = 17$, $S = 0.5$ (f).

Figure 2 (color online). Experimental observation of discrete diffraction, even soliton and three-soliton train. Experimental images at the output crystal face are superimposed on the theoretical plots showing propagation dynamics inside the crystal. Panel (a) shows discrete diffraction in linear regime. Panel (b) shows excitation of an even soliton with two Gaussian beams with amplitudes $A = 5$ and widths $W = 0.5$ launched into adjacent channels, and panel (c) shows excitation of such soliton by single beam with $A = 5$ and $W = 1$ launched in-between channels. (d) Excitation of three-soliton train with broad super-Gaussian beam with $A = 5$ and $W = 2.3$. In all cases $p = 17$ and $S = 0.5$.

Figure 3 (color online). Observed temporal dynamics of the formation of even solitons. (a) Discrete diffraction and (b) temporal build-up of the even soliton for excitation with two in-phase Gaussian beams. In (c) the input beam on the right hand side is blocked at $t = 0\,\text{s}$. Light energy starts to oscillate in the directional-coupler-like waveguiding structure, and finally forms a narrow odd soliton in the excited channel.

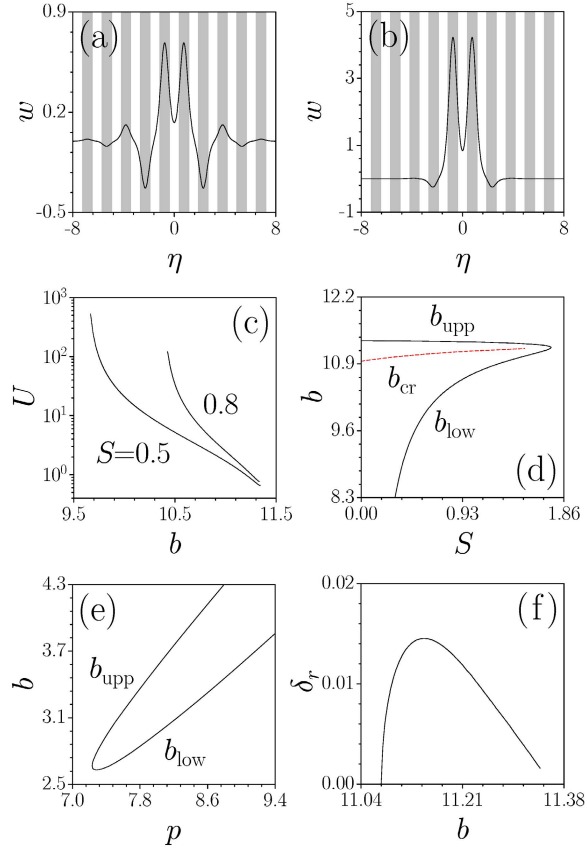

Figure 1 (color online). Profiles of even solitons with $b=11.34$ (a) and $b=10$ (b) at $p=17$ and $S=0.5$. (c) Energy flow versus propagation constant at $p=17$. Domains of existence for even solitons on $(S,b)$ plane at $p=17$ (d) and on $(p,b)$ plane at $S=0.5$ (e). Red (dashed) curve in panel (d) indicates the upper border of stability domain. Real part of perturbation growth rate versus propagation constant at $p=17$, $S=0.5$ (f).

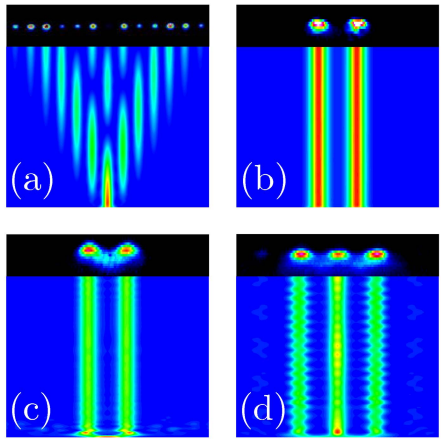

Figure 2 (color online). Experimental observation of discrete diffraction, even soliton and three-soliton train. Experimental images at the output crystal face are superimposed on the theoretical plots showing propagation dynamics inside the crystal. Panel (a) shows discrete diffraction in linear regime. Panel (b) shows excitation of an even soliton with two Gaussian beams with amplitudes $A=5$ and widths $W=0.5$ launched into adjacent channels, and panel (c) shows excitation of such soliton by single beam with $A=5$ and $W=1$ launched in-between channels. (d) Excitation of three-soliton train with broad super-Gaussian beam with $A=5$ and $W=2.3$. In all cases $p=17$ and $S=0.5$.

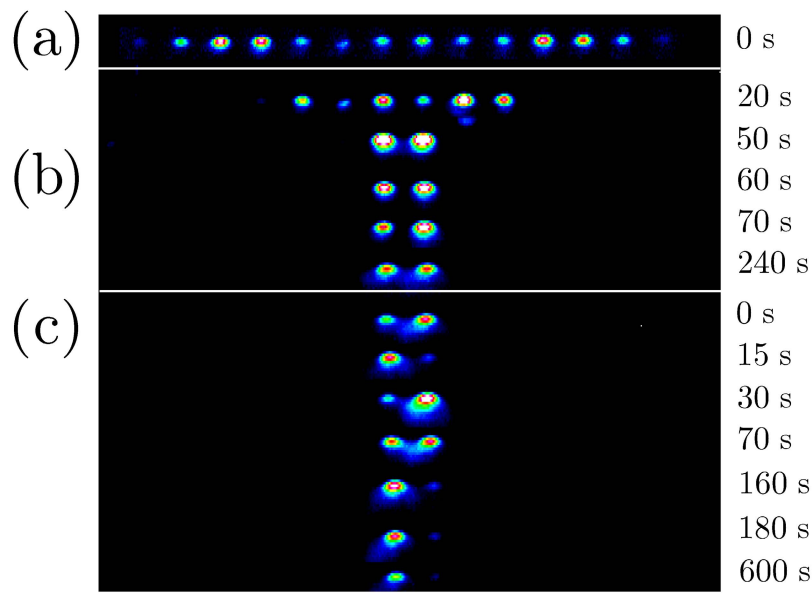

Figure 3 (color online). Observed temporal dynamics of the formation of even solitons. (a) Discrete diffraction and (b) temporal build-up of the even soliton for excitation with two in-phase Gaussian beams. In (c) the input beam on the right hand side is blocked at $t = 0\,\text{s}$. Light energy starts to oscillate in the directional-coupler-like waveguiding structure, and finally forms a narrow odd soliton in the excited channel.

13